\documentclass[12pt]{article}

\usepackage{amsmath,amsfonts,amsbsy,amssymb}
\usepackage{graphicx,psfrag,epsf}
\usepackage{natbib}
\usepackage{algorithm}     
\usepackage{algpseudocode} 
\usepackage[colorlinks=FALSE,citecolor=blue,urlcolor=blue]{hyperref}
\usepackage{bbm}       
\usepackage{booktabs}  
\usepackage[affil-it]{authblk}

\usepackage[english]{babel}
\usepackage[utf8]{inputenc}
\usepackage[T1]{fontenc}
\usepackage{bm,array}

\newcommand{\N}{\mathbb{N}}

\newcommand{\R}{\mathbb{R}}

\newcommand{\E}[1]{\operatorname{\mathbb{E}}[#1]}

\newcommand{\Var}[1]{\operatorname{\mathbb{V}}\left(#1\right)}

\newcommand\independent{\protect\mathpalette{\protect\independenT}{\perp}}
\def\independenT#1#2{\mathrel{\rlap{$#1#2$}\mkern2mu{#1#2}}}

\newcommand{\monthword}[1]{\ifcase#1\or Jan\or Feb\or M\"ar\or Apr\or Mai\or Jun\or Jul\or Aug\or Sep\or Okt\or Nov\or Dez\fi}
\newcommand{\leadingzero}[1]{\ifnum #1<10 0\the#1\else\the#1\fi}             

\newcommand{\todayI}{\the\year"~\leadingzero{\month}"~\leadingzero{\day}}    	
\newcommand{\todayII}{\the\year\leadingzero{\month}\leadingzero{\day}}       	
\newcommand{\todayIII}{\leadingzero{\day}/\leadingzero{\month}/\the\year}    	
\newcommand{\todayIV}{\leadingzero{\day}.\leadingzero{\month}.\the\year}     	
\newcommand{\todayV}{\the\day.\the\month.\the\year}                          	
\newcommand{\todayVI}{\the\day.~\monthword{\the\month}. \the\year}           	
\newcommand{\todayVII}{\leadingzero{\day}.~\monthword{\the\month}. \the\year}	
\newcommand{\todayVIII}{\monthword{\the\month}. \the\year}										

\newcommand{\vd}{\mathbf d}

\newcommand{\vxi}{\bm \xi}

\newcommand{\frakB}{\mathfrak{B}}

\makeatletter
\newcommand{\rd}{\@ifnextchar^{\DIfF}{\DIfF^{}}}
\def\DIfF^#1{%
   \mathop{\mathrm{\mathstrut d}}%
   \nolimits^{#1}\gobblespace}
\def\gobblespace{\futurelet\diffarg\opspace}
\def\opspace{%
   \let\DiffSpace\!%
   \ifx\diffarg(%
   \let\DiffSpace\relax
   \else
   \ifx\diffarg[%
   \let\DiffSpace\relax
   \else
   \ifx\diffarg\{%
   \let\DiffSpace\relax
   \fi\fi\fi\DiffSpace}

\renewcommand{\lim}[1]{\underset{#1}{\operatorname{lim}} \;}

\newcounter{abcd}[section]
\newcommand{\bitabc}{\vspace{-1.8ex}\begin{enumerate}
\renewcommand{\labelenumi}{\alph{abcd})}
\itemsep-1.4ex \partopsep-1.8ex}
\newcommand{\eitabc}{\end{enumerate}}

\newcommand{\bitp}{\vspace{-1ex}\begin{itemize}
\itemsep-0.6ex \partopsep-0.3ex}
\newcommand{\eitp}{\end{itemize}}

\newenvironment{fshaded}{\MakeFramed{\FrameRestore}}{\endMakeFramed}
\newenvironment{orangeumgebungs}[1][]{\definecolor{shadecolor}{rgb}{1,.8,.8}\definecolor{framecolor}{rgb}{1,0,0}\begin{fshaded}\begin{equation*}#1}{\end{equation*}\end{fshaded}}
\newcommand{\bous}{\begin{orangeumgebungs}}
\newcommand{\eous}{\end{orangeumgebungs}}

\newenvironment{blaueumgebungs}[1][]{\definecolor{shadecolor}{rgb}{.9,.9,1}%
\definecolor{framecolor}{rgb}{.1,.0,.7}%
\begin{fshaded}\begin{equation*}#1}{\end{equation*}\end{fshaded}}
\newcommand{\bbus}{\begin{blaueumgebungs}}
\newcommand{\ebus}{\end{blaueumgebungs}}

\newenvironment{blaueumgebunge}[1][]{\definecolor{shadecolor}{rgb}{.9,.9,1}%
\definecolor{framecolor}{rgb}{.1,.0,.7}%
\begin{fshaded}\begin{enumerate}#1}{\end{enumerate}\end{fshaded}}
\newcommand{\bbue}{\begin{blaueumgebunge}}
\newcommand{\ebue}{\end{blaueumgebunge}}

\newcommand{\beq}{\begin{equation}}
\newcommand{\eeq}{\end{equation}}
\newcommand{\beqs}{\begin{equation*}}
\newcommand{\eeqs}{\end{equation*}}

\newcommand{\bal}{\begin{align}}
\newcommand{\eal}{\end{align}}
\newcommand{\bals}{\begin{align*}}
\newcommand{\eals}{\end{align*}}

\newcommand{\bmat}{\begin{pmatrix}}
\newcommand{\emat}{\end{pmatrix}}

\newcommand{\bit}{\begin{itemize}}
\newcommand{\eit}{\end{itemize}}
\newcommand{\biit}{\begin{itemize}}
\newcommand{\eiit}{\end{itemize}}

\newcommand{\ben}{\begin{enumerate}}
\newcommand{\een}{\end{enumerate}}
\newcommand{\been}{\begin{enumerate}}
\newcommand{\eeen}{\end{enumerate}}

\newcommand{\bca}{\begin{cases}}
\newcommand{\eca}{\end{cases}}

\newcommand{\bpa}{\begin{parts}}
\newcommand{\epa}{\end{parts}}

\newcommand{\bspa}{\begin{subparts}}
\newcommand{\espa}{\end{subparts}}

 \newenvironment{NewSolution}
    {\SetTotalwidth\begin{solution}}
    {\end{solution}}
\newcommand{\bso}{\begin{NewSolution}}
\newcommand{\eso}{\end{NewSolution}}
		
\newtheorem{theorem}{Theorem}[section]
\newtheorem{proposition}{Proposition}[section]
\newtheorem{definition}{Definition}[section]

\begin{document}

\title{\bf Partially Observed Functional Data: The Case of Systematically Missing Parts}
\author[a]{Dominik Liebl\thanks{Corresponding author: Dominik Liebl (dliebl@uni-bonn.de), Department of Statistics, University of Bonn, Adenauerallee 24-42, 53113 Bonn, Germany}}
\author[b]{Stefan Rameseder}
\affil[a]{Department of Statistics, University of Bonn}
\affil[b]{Department of Econometrics, University of Regensburg}
\maketitle

\bigskip

\begin{abstract}
New estimators for the mean and the covariance function for partially observed functional data are proposed using a detour via the fundamental theorem of calculus. The new estimators allow for a consistent estimation of the mean and covariance function under specific violations of the missing-completely-at-random assumption. The requirements of the estimation procedure can be tested using a sequential multiple hypothesis test procedure. An extensive simulation study compares the new estimators with the classical estimators from the literature in different missing data scenarios. The proposed methodology is motivated by the practical problem of estimating the mean price curve in the German Control Reserve Market. In this auction market, price curves are only partially observable and the underlying missing data mechanism depends on systematic trading strategies which clearly violate the missing-completely-at-random assumption. In contrast to the classical estimators, the new estimators lead to useful estimates of the mean and covariance functions. Supplementary materials are provided online\footnote{Supplementary materials: \textsf{R}-package \texttt{PartiallyFD} and \textsf{R}-scripts for reproducing the simulation study and the real data application.}.
\end{abstract}


\newpage

\section{Introduction}\label{sec:int}
The classical literature on Functional Data Analysis (FDA) focuses on the analysis of functions where each function $X_i(t)$ is observable for all $t\in[a,b]$ (see, for instance, the textbooks \citeauthor{RamSil2005}, \citeyear{RamSil2005}, \citeauthor{FerVie2006}, \citeyear{FerVie2006}, \citeauthor{HorKok2012}, \citeyear{HorKok2012}, and \citeauthor{HsiEub2015}, \citeyear{HsiEub2015}). However, this regular situation does not apply to many functional data sets of practical relevance where the functions $X_i$ are only partially observable, i.e., where $X_i(t)$ is only observable for $t\in D_i$ with $D_i\subset[a,b]$ describing the $i$-specific observable subset of the total domain.

The latter situation is often referred to as fragmented, truncated, incomplete, or partially observed functional data and its practical relevance has triggered a series of research works dealing with different aspects of this problem. \citet{DelHal2013} propose a ``shift-and-connect'' procedure to reconstruct and classify fragmentary functional data and \citet{DelHal2016} incorporate a Markov chain model. \cite{ZhoSerGebMue2014} model truncated functional data using warping functions. \cite{GolRitMan2014} and \citet{Kra2015} use functional linear regression models to predict the missing parts. \cite{GelColNeeCrai2014} and \cite{GroKokSoj2017} propose a functional regression model for incomplete curves. \cite{Lie2013} models and forecasts partially observed price functions and \citet{LieKne2018} consider optimal reconstructions of partially observed functional data. All these works, however, make use of the so-called Missing-Completely-At-Random (MCAR) assumption, i.e., the assumption that the missing data mechanism is independent from all other stochastic components of relevance \citep[see, for instance,][Ch.~1]{LitRub2014}. All of the above cited works lead to inconsistent results if the MCAR assumption is violated. To the best of our knowledge, we are the first to consider specific violations of the MCAR assumption in the context of partially observed functional data.

The classical missing data literature can be divided into the following four---not mutually exclusive---classes \citep[see][Ch.~1]{LitRub2014}: first, procedures that discard all incompletely recorded units on the premise that the MCAR assumption holds; second, procedures that use a re-weight\-ing of the data in order to adjust for the missing data; third, procedures that impute missing values using simple estimators (e.g., mean imputation); and fourth, procedures that are based on model assumptions. In this work, we add a fifth class that exclusively applies to functional data due to its general availability of derivatives. By using a detour via the fundamental theorem of calculus, we propose an estimation procedure that allows the consistent estimation of the mean and covariance function under specific violations of the MCAR assumption. In order to test for the considered violations of the MCAR assumption, we propose the application of the sequential multiple hypothesis testing procedure of \citet{RomWol2005b}.

Note that we focus on the regular case of functional data where the single functions are only partially observed, but where the observed parts are fully observed. That is, we do not consider the case of sparse functional data where one observes only a few real, possibly noise contaminated, discretization points per function (see \citeauthor{JamHasSu2000}, \citeyear{JamHasSu2000}, \citeauthor{JamSug2003}, \citeyear{JamSug2003}, \citeauthor{YaoMueWa2005}, \citeyear{YaoMueWa2005}, and \citeauthor{YaoMueWan2005}, \citeyear{YaoMueWan2005}, among others). Our estimation procedure requires the availability of derivatives and, therefore, cannot directly deal with the case of sparse functional data.

Our work is motivated by a real data set from energy economics with partially observed price curves $X_i(t)$. For each $X_i(t)$ we only observe the initial part for all $t\in[a,d_i]$; however, the final part with $t\in(d_i,b]$ is missing. The specific market structure incentivizes a systematic bidding strategy that results in a missing data mechanism where the random variable $d_i$ correlates with the overall level of the price curves $X_i$. Larger values of $d_i$ are associated with price curves having an overall high price level and vice versa (see Figure \ref{fig:ftc_mean2}). The general exposition of our methodology is geared towards the data situation in this real data application, which provides a simple and instructive walk-along setup. However, we generalize this introducing setup also for broader violations of the MCAR assumption and broader missing data designs.

The rest of the paper is structured as follows. The next section introduces the setup under consideration, our statistical methodology, and generalizations. In Section \ref{sec:vio}, we propose a practical approach for testing the considered violations of the MCAR assumption. Section \ref{sec:sim} contains our simulation study and the application is found in Section \ref{sec:emp}. Proofs and further derivations can be found in Appendix \ref{sec:app}.

\section{Methodology} \label{sec:met}
\subsection{General Setup}\label{ssec:set}
We consider an i.i.d.~sample of differentiable random functions $X_1, \ldots, X_n$ each with the same distribution as $X$ with values in the separable Hilbert space $L^2([a,b])$, where we set $[a,b]=[0,1]\subset\R$ without loss of generality. We assume that $\E{||X||_2^4}<\infty$, where $||X||_2^2=\int_0^1X(t)^2dt$. The mean and covariance functions are denoted by $\mu(t)=\E{X(t)}$ and $\sigma(s,t)=\E{(X(s)-\mu(s))(X(t)-\mu(t))}$.

Motivated by our real data application we consider the following missing data mechanism: the random functions $X_i$ are only observable over random subdomains $D_i=[0, d_i] \subseteq [0,1]$. Here, $d_i$ are i.i.d.~copies of a real random variable $d$ with density $f_d$, where $f_d(t)>0$ for $t\in[d_{\operatorname{min}},1]$ and zero else, with deterministic $0<d_{\operatorname{min}}<1$. The random subdomains $D_i=[0, d_i]$ lead to a $t$-specific observed data indicator $O_i(t)$ defined as $O_i(t)=\mathbbm{1}_{t\in D_i}$; i.e., $O_i(t)=1$ if $X_i(t)$ is observable and $O_i(t)=0$ if $X_i(t)$ is missing. Let $p(t)=\mathbb{P}(O_i(t)=1)$ denote the probability of observing functions covering $t$. Observe that under our setup $p(t)=1$ for all $t\in[0,d_{\operatorname{min}}]$ and $0<p(t)<1$ for all $t\in(d_{\operatorname{min}},1]$. That is, the random functions do not contain missing parts over the lower interval $[0,d_{\operatorname{min}}]$, but may have missing endings over the upper interval $(d_{\operatorname{min}},1]$.

The MCAR assumption requires independence between the entire random processes $O_i$ and $X_i$. Under our setup, however, a violation of the MCAR assumption can only affect the upper interval $(d_{\operatorname{min}},1]$, i.e., the part of the domain where the functions may have missing endings. That is, any violation of the MCAR assumption is without effect in the lower interval $[0,d_{\operatorname{min}}]$, where the functions do not contain missing observations; see Figure \ref{fig:ftc_mean2} for a real data example.

Our estimators use a detour via the fundamental theorem of calculus which allows us to address some practically relevant violations of the MCAR assumption. To formalize these violations, further notation needs to be introduced. Decompose $X(t) = \sum_{j\geq 1} \xi_{j} \psi_j(t)$, where $\frakB=\lbrace \psi_1,\psi_2,\dots\rbrace$ forms a deterministic orthogonal basis system of $L^2([0,1])$ with $\psi_1 \equiv 1$, and where $\xi_{j}=\int_0^1X(t)\psi_j(t)dt$, with $j\geq 1$, denote the---not necessarily centered---random basis coefficients. Common basis systems which fulfill these assumptions are, for instance, a Fourier basis system or Legendre polynomials after applying an appropriate scaling of the domain. Denoting $S=\operatorname{span}\{1\}$ and $S^\perp=\operatorname{span}\{\psi_2,\psi_3,\dots\}$ allows to separate the functions $X=X^S+X^{S^\perp}$, where  $X^S$ and $X^{S^\perp}$ are the orthogonal projections of $X$ on $S$ and $S^\perp$.
For a start, we consider the following Violation (V) of the MCAR assumption: 
\begin{itemize}
\item[(V)] $X\not\independent O$, but $X^{S^\perp}\independent O$; i.e, the dependency between $X$ and $O$ manifests only in the dependency between $O$ and the vertical shift component $\xi_{1}\psi_1=\xi_{1}$ of $X$. 
\end{itemize}

Note that Violation (V) causes distortions of the mean and covariance functions for every $t\in(d_{\operatorname{min}},1]$, since $\mathbb{E}[\xi_1|O(t)]\neq\mathbb{E}[\xi_1]$ for all $t\in(d_{\operatorname{min}},1]$. For $t\in[0,d_{\operatorname{min}}]$, however, Violation (V) is ineffective, since $O(t)=1$ almost surely, such that $\mathbb{E}[\xi_j|O(t)]=\mathbb{E}[\xi_j]$ for all $j\geq 1$ and all $t\in[0,d_{\operatorname{min}}]$.

This relatively simple Violation (V) of the MCAR assumption is motivated by our real data application and allows for a comprehensible and pithy explanation of our estimation strategies. Below, in Section \ref{subsec:gen}, we introduce more general versions of Violation (V) and discuss how to apply our estimation strategies accordingly.

\subsection{FTC-Estimators}
In the case of partially observed functional data, it is impossible to use the classical mean estimator, $\bar{X}(t)=n^{-1}\sum_{i=1}^nX_i(t)$, and covariance estimator, $n^{-1}\sum_{i=1}^n(X_i(t)-\bar{X}(t))(X_i(s)-\bar{X}(s))$, since the value of $X_i(t)$ may not be observed. Therefore, \cite{DelHal2013} and \citet{Kra2015} propose the use of the following estimators:
\begin{align}
\widehat\mu(t) &= \frac{I_1(t)}{\sum_{i=1}^n O_i(t)} \sum_{i=1}^n X_i(t)O_i(t),\label{krauss}\\
\widehat\sigma(s,t) &= \frac{I_2(s,t)}{\sum_{i=1}^n U_i(s,t)} \sum_{i=1}^n U_i(s,t) \left[X_i(s) - \widehat\mu(s) \right] \left[X_i(t) - \widehat\mu(t) \right],\label{var_kra}
\end{align}
where the existence functions $I_1(t) = \mathbbm{1}_{\sum_{i=1}^n O_i(t) > 0}$ and $I_2(s,t)=\mathbbm{1}_{\sum_{i=1}^n U_i(s,t) > 0}$ for $U_i(s,t) = O_i(s)O_i(t)$ are necessary to prevent divisions by zero through defining $0/0=\text{NA}$.

Violation (V) of the MCAR assumption implies that the estimator $\widehat\mu(t)$ is biased and inconsistent for all $t\in(d_{\operatorname{min}},1]$, i.e., for all $t$ with $p(t)<1$. The same applies to the estimator $\widehat\sigma(s,t)$ if at least one of the arguments $s$ or $t$ is an element of the critical set $(d_{\operatorname{min}},1]$. Simple rearrangements using the law of iterated expectations show that under our setup
\begin{align*}
\E{\widehat\mu(t)} = \mu(t) + \Delta(t),\quad\text{where}\quad\Delta(t)=\mathbb{E}\left[\frac{I_1(t)}{\sum_{i=1}^n O_i(t)} \sum_{i=1}^n O_i(t)\mathbb{E}\left[\xi_{i1}\hspace{1mm}|\hspace{1mm}O_i(t)\right]\right] - \mu_1,
\end{align*}
with $|\Delta(t)|>0$, i.e., $\mathbb{E}\left[\xi_{i1}\hspace{1mm}|\hspace{1mm}O_i(t)\right]\neq\mathbb{E}\left[\xi_{i1}\right]$, iff $p(t)<1$; an essentially equivalent bias expression applies to $\widehat\sigma(s,t)$ (see also Proposition \ref{lem:bias} in Appendix \ref{sec:app}).

Therefore, we propose the following detour via the Fundamental Theorem of Calculus (FTC) which leads to consistent estimators of the mean and covariance functions under Violation (V) of the MCAR assumption. The FTC states that one can decompose a differentiable function $f(t)$ as $f(t) = \int_0^t f^{(1)}(z) dz + f(0)$, where $f^{(1)}$ denotes the first derivative of $f$. This motivates our new estimators $\widehat\mu_{\text{FTC}}(t)$ and $\widehat\sigma_{\text{FTC}}(s,t)$ for the mean and covariance functions.
\begin{definition}[FTC-mean]
The FTC mean estimator is defined as
\beq\label{mean_ftc}
\widehat\mu_{\operatorname{FTC}}(t) =
\begin{cases}
\widehat\mu(t) & \text{if}\quad t \in [0, d_{\operatorname{min}}]\\
\int_{d_{\operatorname{min}}}^t \widehat\mu^{(1)}(z) dz + \widehat\mu(d_{\operatorname{min}})    & \text{if}\quad t \in (d_{\operatorname{min}}, 1],
\end{cases}
\eeq
where 
\begin{align*}
\widehat\mu^{(k)}(t)= \frac{I_1(t)}{\sum_{i=1}^n O_i(t)} \sum_{i=1}^n X^{(k)}_i(t)O_i(t),
\end{align*}
and where $X^{(k)}_i(t)$ denotes the $k$-th derivative of $X_i(t)$.
\end{definition}

\begin{definition}[FTC-covariance]\label{def:FTC_cov}
The FTC covariance estimator is defined as
\beq \label{var_ftc}
\widehat{\sigma}_{\operatorname{FTC}}(s,t) =
\begin{cases}
\widehat{\sigma}(s,t) & \text{if } (s,t)\in[0, d_{\operatorname{min}}]^2 \\[1.5ex]
\int_{d_{\operatorname{min}}}^t \widehat{\sigma}^{(0,1)}(s, z_2)dz_2 + \widehat{ \sigma}(s,d_{\operatorname{min}}) & \text{if } (s,t)\in [0, d_{\operatorname{min}}] \times (d_{\operatorname{min}}, 1] \\[1.5ex]
\int_{d_{\operatorname{min}}}^s \widehat{\sigma}^{(1,0)}(z_1,t)dz_1 + \widehat{ \sigma}(d_{\operatorname{min}},t) & \text{if } (s,t)\in (d_{\operatorname{min}}, 1]\times [0, d_{\operatorname{min}}]\\[1.5ex]
\int_{d_{\operatorname{min}}}^t \int_{d_{\operatorname{min}}}^s \widehat{\sigma}^{(1,1)}(z_1,z_2) dz_1dz_2 + \\
\int_{d_{\operatorname{min}}}^s \widehat{\sigma}^{(1,0)}(z_1,d_{\operatorname{min}}) dz_1 +\\
\int_{d_{\operatorname{min}}}^t \widehat{\sigma}^{(0,1)}(d_{\operatorname{min}},z_2) dz_2 +
\widehat{\sigma}(d_{\operatorname{min}},d_{\operatorname{min}}) & \text{if } (s,t) \in (d_{\operatorname{min}}, 1]^2,
\end{cases}
\eeq
where 
\begin{align*}
\widehat{\sigma}^{(\ell,k)}(s,t)=
\frac{I_2(s,t)}{\sum_{i=1}^n U_i(s,t)} \sum_{i=1}^n U_i(s,t) \left[X_i^{(\ell)}(s) - \widehat{\mu}^{(\ell)}(s) \right] \left[X_i^{(k)}(t) - \widehat{\mu}^{(k)}(t) \right],\quad \ell,k\in\{0,1\}.
\end{align*}
\end{definition}

\smallskip

A step-by-step derivation of $\widehat \sigma_{\operatorname{FTC}}(s,t)$ can be found in parts I-III of the proof of Theorem \ref{Th1} (see Appendix \ref{sec:app}). The following theorem states the consistency of the FTC-estimators under Violation (V) of the MCAR assumption:
\begin{theorem}\label{Th1}
Under our setup and under Violation $(V)$ of the MCAR assumption, the estimators $\widehat{\mu}_{\operatorname{FTC}}(t)$, defined in \eqref{mean_ftc}, and $\widehat{\sigma}_{\operatorname{FTC}}(s,t)$, defined in \eqref{var_ftc}, are pointwise $\sqrt{n}$-consistent estimators of $\mu(t)$ and $\sigma(s,t)$ for all $s,t\in[0,1]$.
\end{theorem}

\subsection{Generalizations}\label{subsec:gen}
Motivated by our real data application, we considered so far a relatively simple missing data design that can be characterized by the following two restrictions: first, $O_i(t)$ only depends on vertical shifts in $X_i(t)$, and second, the domain $[0,1]$ is divided into two subdomains $[0, d_{\operatorname{min}}]$ and $(d_{\operatorname{min}},1]$, where the first is covered by all functions, i.e., $p(t)=1$ for all $t\in[0, d_{\operatorname{min}}]$, and where the second contains the missing parts, i.e., $0<p(t)<1$ for all $t\in(d_{\operatorname{min}},1]$. In the following, we generalize both of these restrictions.

The latter domain restriction can be weakened considerably, since the FTC estimators need only one point where the full sample of curves is observed. Denote this point by $d_{\text{f}} \in [0,1]$, i.e., $p(d_{\text{f}})=1$. By noting $\int_a^b = -\int_b^a$, the generalized FTC mean estimator  is defined as
\beq\label{genmean_ftc}
\widehat\mu^\star_{\operatorname{FTC}}(t) =
\begin{cases}
\phantom{-}\int_{d_{\text{f}}}^t \widehat\mu^{(1)}(z) dz + \widehat\mu(d_{\text{f}})&\text{for all $t>d_{\text{f}}$}\\[1.5ex]
\phantom{-}\widehat{\mu}(d_{\text{f}})&\text{for $t=d_{\text{f}}$}\\[1.5ex]
-\int^{d_{\text{f}}}_t \widehat\mu^{(1)}(z) dz + \widehat\mu(d_{\text{f}})&\text{for all $t<d_{\text{f}}$}.

\end{cases}
\eeq

Correspondingly, the generalized FTC covariance estimator is defined as
\begin{align}\label{genvar_ftc}
&\widehat{\sigma}^\star_{\operatorname{FTC}}(s,t) =\nonumber\\
&=\int_{d_{\text{f}}}^t \int_{d_{\text{f}}}^s \widehat{\sigma}^{(1,1)}(z_1,z_2) dz_1dz_2 + 
\int_{d_{\text{f}}}^s \widehat{\sigma}^{(1,0)}(z_1,d_{\text{f}}) dz_1 +
\int_{d_{\text{f}}}^t \widehat{\sigma}^{(0,1)}(d_{\text{f}},z_2) dz_2 +
\widehat{\sigma}(d_{\text{f}},d_{\text{f}})
\end{align}
for all $s,t>d_{\text{f}}$, where for cases with $s<d_{\text{f}}$ and/or $t<d_{\text{f}}$ one needs to replace ``$\int_{d_{\text{f}}}^s$'' with ``$-\int^{d_{\text{f}}}_s$'' and/or ``$\int_{d_{\text{f}}}^t$'' by ``$-\int^{d_{\text{f}}}_t$'', and where $\widehat{\sigma}^\star_{\operatorname{FTC}}(s,t)=\widehat{\sigma}(d_{\text{f}},d_{\text{f}})$ for $s=t=d_{\text{f}}$. The estimators in \eqref{genmean_ftc} and \eqref{genvar_ftc} can be trivially adjusted for scenarios when the observation mechanism contains larger fragments of fully observed samples.

In order to generalize Violation (V), define $S_K=\operatorname{span}\{\psi_1,\dots,\psi_K\}$ with monomial basis functions $\psi_j(t)=t^{j-1}$ for $j=1,\dots,K$ and let $S_K^{\perp}$ denote the orthogonal complement of $S_K$ in $L^2([0,1])$. Additionally, we need to assume that $X$ is $K$-times differentiable. The following violation of the MCAR assumption generalizes Violation (V): 
\begin{itemize}
\item[(V$^K$)] $X\not\independent O$, but $X^{S_K^\perp}\independent O$; i.e, the dependency between $X$ and $O$ manifests only in the dependency between $O$ and the first $K$ monomial components $\xi_{1}\psi_1,\dots,\xi_{K}\psi_K$ of $X$. 
\end{itemize}

Our estimation procedure can be applied to this more general Violation (V$^K$) of the MCAR assumption using the following recursion, which back-transforms the consistent estimator $\widehat{\mu}^{(K)}$ by repeatedly applying the FTC:
\begin{align*}
\widetilde{\mu}^{(K-1)}_{\text{FTC}}(t)&=\int_{d_{\text{f}}}^t\widehat{\mu}^{(K)}(z)dz+\widehat\mu^{(K-1)}(d_{\text{f}}),\\
\widetilde{\mu}^{(K-2)}_{\text{FTC}}(t)&=\int_{d_{\text{f}}}^t\widetilde{\mu}^{(K-1)}_{\text{FTC}}(z)dz+\widehat\mu^{(K-2)}(d_{\text{f}}),\\
\vdots\quad\quad &\\
\text{and}\quad\widetilde{\mu}_{\text{FTC}}(t)&=\int_{d_{\text{f}}}^t\widetilde{\mu}^{(1)}_{\text{FTC}}(z)dz+\widehat\mu(d_{\text{f}}),
\end{align*}
where $\widetilde\mu_{\text{FTC}}(t)$ denotes the (back-transformed) estimator of $\mu(t)$. 

An equivalent, yet more tedious recursion can be applied in order to back-transform the consistent estimator $\widehat{\sigma}^{(K,K)}$:
\begin{align*}
\widetilde{\sigma}^{(K-1, K-1)}_{\text{FTC}}(s,t)
&= \int_{d_{\text{f}}}^t \int_{d_{\text{f}}}^s \widehat{\sigma}^{(K,K)}(z_1,z_2) dz_1dz_2 + \int_{d_{\text{f}}}^s \widehat{\sigma}^{(K,K-1)}(z_1,d_{\text{f}}) dz_1 + \\
&+ \int_{d_{\text{f}}}^t \widehat{\sigma}^{(K-1,K)}(d_{\text{f}},z_2) dz_2 + \widehat{\sigma}^{(K-1, K-1)}(d_{\text{f}},d_{\text{f}}),\\
\widetilde{\sigma}^{(K-2, K-2)}_{\text{FTC}}(s,t)
&= \int_{d_{\text{f}}}^t \int_{d_{\text{f}}}^s \widetilde{\sigma}^{(K-1,K-1)}_{\text{FTC}}(z_1,z_2) dz_1dz_2 + 
  \int_{d_{\text{f}}}^s \int_{d_{\text{f}}}^{t}\widehat{\sigma}^{(K,K-2)}(z_1,d_{\text{f}}) dz_1 dz_1\\
&+\int_{d_{\text{f}}}^t \int_{d_{\text{f}}}^{t}\widehat{\sigma}^{(K-2,K)}(d_{\text{f}},z_2) dz_2dz_2 + 2  \widehat{\sigma}^{(K-1,K-2)}(d_{\text{f}},d_{\text{f}}) + \widehat{\sigma}^{(K-2, K-2)}(d_{\text{f}},d_{\text{f}}),\\
\vdots\quad\quad &\\
\text{and}\quad\widetilde{\sigma}_{\text{FTC}}(s,t)
&= \int_{d_{\text{f}}}^t \int_{d_{\text{f}}}^s \widetilde{\sigma}^{(1,1)}_{\text{FTC}}(z_1,z_2) dz_1dz_2 + 
 \int_{d_{\text{f}}}^s \int_{d_{\text{f}}}^{t} \widehat{\sigma}^{(0,1)}(z_1,d_{\text{f}}) dz_1 dz_1\\
&+ \int_{d_{\text{f}}}^t \int_{d_{\text{f}}}^{t} \widehat{\sigma}^{(0,1)}(d_{\text{f}},z_2) dz_2dz_2 + 2  \widehat{\sigma}^{(1,0)}(d_{\text{f}},d_{\text{f}}) + \widehat{\sigma}^{(0, 0)}(d_{\text{f}},d_{\text{f}}),
\end{align*}
where $\widetilde{\sigma}_{\text{FTC}}(s,t)$ denotes the (back-transformed) estimator of $\sigma_{\text{FTC}}(s,t)$.

\section{A Practical Approach to Test Violation (V)} \label{sec:vio}
In practice, there are situations where Violation (V) is essentially known from the context such as, for instance, in our real data application. Generally, however, practitioners would like to test for Violation (V) in order to gain confidence for applying our FTC-estimators. In the following, we focus on testing Violation (V); however, the testing procedure can be easily generalized to the more general Violation (V$^K$) as well.

Unfortunately, a proper test for Violation (V) cannot---to the best of our knowledge---be achieved using existing procedures. A proper test procedure needs to investigate possible dependencies between $O_i$ and the coefficients $\xi_{i1},\xi_{i2},\dots$. The coefficients $\xi_{i1},\xi_{i2},\dots$, however, refer to the complete trajectories and, therefore, cannot be computed from the partially observed trajectories. Furthermore, existing procedures that allow to predict the coefficients $\xi_{i1},\xi_{i2},\dots$ from the partially observed trajectories require the possibly violated MCAR assumption (see, for instance, \citeauthor{Kra2015}, \citeyear{Kra2015}, or \citeauthor{YaoMueWa2005}, \citeyear{YaoMueWa2005}).

Therefore, we can only propose a practical test procedure which will be useful for well-structured functional data, but generally cannot be applied in the case of complex structured functional data. The idea is to use the feasible coefficients $\xi^{[0,d_{\operatorname{min}}]}_{i1},\xi^{[0,d_{\operatorname{min}}]}_{i2},\dots$ with respect to the sub-part of the domain $[0,d_{\operatorname{min}}]$ over which the full sample of functions is observed. For simply structured functional data, the coefficients $\xi^{[0,d_{\operatorname{min}}]}_{i1},\xi^{[0,d_{\operatorname{min}}]}_{i2},\dots$ will be informative about the infeasible coefficients $\xi_{i1},\xi_{i2},\dots$. In fact, \citet{LieKne2018} argue in a related context that the coefficients $\xi^{[0,d_{\operatorname{min}}]}_{i1},\xi^{[0,d_{\operatorname{min}}]}_{i2},\dots$ contain the same information as $\xi_{i1},\xi_{i2},\dots$ if the functional data are such that equality over $[0,d_{\operatorname{min}}]$ implies equality over the total domain $[0,1]$; i.e., if $X_i(t)=X_j(t)$ for all $t\in[0,d_{\operatorname{min}}]$ and $i\neq j$ implies that $X_i(t)=X_j(t)$ for all $t\in[0,1]$. The latter might hold for simply structured functional data and is fulfilled, for instance, for finite dimensional random functions $X_i(t)=\sum_{j=1}^K\xi_{ij}\psi_k(t)$, as long as the basis functions $\psi_1,\dots,\psi_K$ are linearly independent over $[0,d_{\operatorname{min}}]$.

To identify cases falling under Violation (V), we propose the use of the following simple, practical procedure. Assuming Gaussian random functions $X_i$, it suffices to consider correlations in order to check the independence assumption in Violation (V). Under Violation (V), the upper threshold $d_i$ of the partially observed domain $[0,d_i]$ is correlated with $\xi_{i1}$, but mutually uncorrelated with $\xi_{ij}$ for all $j\geq 2$. In order to detect such a correlation structure, we propose the projection of the commonly observed parts of the random functions $X_i(t)$ for all $t\in[0,d_{\operatorname{min}}]$ onto a $J<\infty$ dimensional Fourier basis system, where the number of basis functions $J$ is selected using the Bayesian Information Criterion (BIC): first, we select BIC-optimal numbers of basis functions $J^{\text{BIC}}_i$ for all $i=1,\dots,n$, and second, we use $J=\operatorname{median}\{J^{\text{BIC}}_1,\dots,J^{\text{BIC}}_n\}$ as an overall trade-off between over- and underspecification.

Projecting the random functions $X_i(t)$ for all $t\in[0,d_{\operatorname{min}}]$ onto a $J<\infty$ dimensional Fourier basis system allows us to approximate the feasible coefficients $\xi^{[0,d_{\operatorname{min}}]}_{i1},\dots,\xi^{[0,d_{\operatorname{min}}]}_{iJ}$. In order to test for possible correlations between $\xi^{[0,d_{\operatorname{min}}]}_{i1},\dots,\xi^{[0,d_{\operatorname{min}}]}_{iJ}$ and $d_i$, we use the bootstrap-based multiple testing of \citet{RomWol2005b} which allows to control for the multiple comparisons problem involved.

The procedure of \citet{RomWol2005b} controls the Family-Wise Error Rate (FWER) under an arbitrary set of dependence structures among the test statistics. Furthermore, their procedure allows for a so-called strong control of the FWER, i.e., the false rejection rate $\alpha$ is guaranteed for any combination of true and non-true null hypotheses in opposite to weak control, where a false rejection rate is only guaranteed if all null hypotheses are true \citep[see][for more details]{RomWol2005b}.

We apply their sequential testing procedure to the multiple linear regression model $d_i = \beta_0 + \sum_{j=1}^J \beta_j \xi_{ji} + u_i$ (see Algorithm \ref{alg:rom}), where we substitute the $\xi_{ji}$ with the approximated versions of the $\xi^{[0,d_{\operatorname{min}}]}_{ij}$. We investigate the set of hypotheses $B=\lbrace \text{H}_{0,1}, \ldots, \text{H}_{0,J}\rbrace$ with $\text{H}_{0,j}: \beta_j = 0$, $j = 1, \ldots, J$, using standard squared $t$-test statistics $\hat t_j^2= (\widehat \beta_j / \hat s_{\widehat \beta_j})^2$. Although there are a total of $2^J$ possibly different test outcomes, we are only interested in the following three different types of outcomes:
\bit
\item[Null:] The overall null: all $H_{j,0}, j = 1, \ldots, J,$ cannot be rejected, i.e., all parameters $\beta_j$ are insignificant and their basis coefficients $\xi_{ij}$ do not correlate with $d_i$. This test decision point towards the MCAR assumption.
\item[(V):] Only $H_{1,0}: \beta_1 = 0$ will be rejected, all other $H_{j,0}, j =2, \ldots, J,$ cannot be rejected. That is, only $\xi_{i1}$ significantly correlates with $d_i$. This test decision point towards Violation (V) of the MCAR assumption. 
\item[Other] Any of the remaining $(2^J-2)$ different test decisions. These test decisions point towards more complicated violations of the MCAR assumption.
\eit
Given $n$ observations, $J$ hypotheses, and a significance level $\alpha$, the procedure proposed by \citet{RomWol2005b} is basically to bootstrap the statistics and calculate the bootstrapped $p$-value. For our special case, we can use a simplified version of the original algorithm \citep[see][Algorithm 1 and Equation (26)]{RomWol2005b}. Our algorithm (see Algorithm \ref{alg:rom}) is initialized by setting $B=B_1$, where $\vd = (d_1, \ldots, d_n)$, $\mathbf \Xi = (\vxi_1, \ldots, \vxi_J) \in \R^{n \times J}$, and $R$ denotes the number of bootstrap replications: 
\begin{algorithm}[!ht]
\caption{Romano Wolf Stepdown Procedure for Multiple Hypothesis Testing.}\label{alg:rom}
\begin{algorithmic}[1]
\Procedure{RomanoWolf}{$\vd, \mathbf \Xi, \alpha, R$}
\State Calculate the test statistics $\hat t_j^2, j = 1, \ldots, J,$ for the sample $(\vd, \mathbf \Xi)$.
\State Perform a model-based bootstrap to obtain statistics $\hat t^2_{j,r}, j = 1, \ldots, J, r=1, \ldots, R$.
\For{$j = 1, \ldots, J$ hypotheses}
\State Define $\hat t^2_{b_j}$ the maximum of all (remaining) test statistics in $B_j$ with $b_j = \# B_j$.
\State Define the $p$-value for $b_j$ hypotheses as
\beqs
\widehat{p}_{R, b_j} = \frac{1}{R}\left[1+\sum_{r=1}^{R-1} \mathbbm{1}\lbrace \hat t^2_{j,r} \geq \hat t^2_j\rbrace \right]
\eeqs
\If{$\widehat{p}_{R, b_j} > \alpha$}
        \State stop and accept all remaining hypotheses in $B_j$,
\Else
			\State reject the $b_j$-th hypothesis and define a new set of nulls $B_{j+1} = B_{j}\setminus H_{0, b_j}$.
\EndIf
\EndFor
\State \textbf{return} $B = B_{j}$ the final set of hypotheses and $p=\widehat{p}_{R, b_j}$ the $p$-value.
\EndProcedure
\end{algorithmic}
\end{algorithm}

A further simple visual tool for checking Violation (V)  is to compare the estimates of the classical estimators, $\widehat{\mu}(t)$ and $\widehat{\sigma}(s,t)$, with those of the FTC estimators, $\widehat{\mu}_{\text{FTC}}(t)$ and $\widehat{\sigma}_{\text{FTC}}(s,t)$. Under the MCAR assumption the two estimates of the mean function and the two estimates of the covariance function will essentially coincide, however, under Violation (V) the pairs of estimates will differ.

\section{Simulation} \label{sec:sim}
In the following simulation study, we assess the finite sample properties of our FTC-estimators $\widehat\mu_{\operatorname{FTC}}$ and $\widehat\sigma_{\operatorname{FTC}}$, and investigate performance of the multiple testing procedure described in Algorithm \ref{alg:rom}. For a feasible data generation $X_i(t) = \sum_{j \geq 1} \xi_{ij}\psi_j(t)$, we use a finite dimensional Fourier basis $\psi_1(t) = 1, \psi_{2k}(t) = \sqrt{2} \sin(2\pi k t)$, and $\psi_{2k+1}(t) = \sqrt{2} \cos(2\pi k t)$, $k = 1, \ldots, (J-1)/2$, with $J=5$ and draw uncorrelated coefficients $\xi_{ij}, j = 1, \ldots, J,$ from a normal distribution with means $\E{\xi_{ij}}=\mu_j$ and variances $\Var{\xi_{ij}}=\lambda_j$, where $\mu_1 = 5, \mu_2=2$, $\mu_3 = 0, \dots,\mu_5=0$, and $\lambda_1 = 10, \lambda_2 = 8, \ldots, \lambda_5 = 2$. This results in the mean function $\mu(t) = 5 + 2\cdot\sqrt{2}\cdot\sin(2\pi t)$ and the covariance function $\sigma(s,t) = \sum_{j=1}^J \lambda_j \psi_j(s)\psi_j(t)$.

We consider four different Data Generating Processes (DGPs), as described in \autoref{tab:dgp}. The DGPs represent discrete (Dis.), i.e., $d_i\in \lbrace 0.5, 1 \rbrace$, and continuous (Con.), i.e., $d_i\in [0.5, 1]$, distributions for $d_i$, as well as dependent (Dep.), i.e., $\xi_{i1}\not\independent d_i$, and independent (Ind.), i.e., $\xi_{i1}\independent d_i$, setups. The Dep.-DGPs represent versions of Violation (V) of the MCAR assumption and the Ind.-DGPs fulfill the MCAR assumption.
\begin{table}[!htbp]
\centering
\caption{Data Generating Processes.}
 \begin{tabular}{lcccc}
\toprule 
  & \text{Dep.-Dis.} & \text{Dep.-Con.} & \text{Ind.-Dis.} & \text{Ind.-Con.}\\
\midrule
 $\xi_{i1}$ & $\text{N}(\mu_1, \lambda_1)$ &	$\text{N}(\mu_1, \lambda_1)$& $\text{N}(\mu_1, \lambda_1)$ &  $\text{N}(\mu_1, \lambda_1)$\\
 $d_i$ 	& $\text{sgn}_{\lbrace0.5,1\rbrace}(\xi_{i1}^c)$		&	$\text{Unif}^{\text{sc}}(\xi_{i1}; \mu_1, \lambda_1)$	& $\text{Ber}_{\lbrace 0.5,1\rbrace}(0.5)$	&  $\text{Unif}([0.5, 1])$\\
\bottomrule
 \end{tabular}
 \label{tab:dgp}
 \end{table}

In each of the DGPs, $\xi_{i1}$ is drawn from N$(\mu_1, \lambda_1)$. For the Dep.-Dis.~DGP, we define $d_i=0.5$ if $\xi_{i1}^c=\xi_{i1}-\mu_1<0$ and $d_i=1$ if $\xi_{i1}^c>0$. In the Dep.-Con.~DGP, we define $d_i$ as a transformed version of $\xi_{i1}$ using the following two steps: first, we use the uniform integral transformation to transform $\xi_{i1}$ to a uniformly (on $[0,1]$) distributed random variable $\Phi_{\mu_1,\lambda_1}(\xi_{i1})$, where $\Phi_{\mu_1,\lambda_1}$ denotes the distribution function of the Normal distribution with parameters $\mu_1$ and $\lambda_1$. Second, we project all values of $\Phi_{\mu_1,\lambda_1}(\xi_{i1})$ smaller 0.5 onto 0.5 and force the 2 percent largest values onto 1, which results in a mixture (discrete and continuous) distribution with point masses at 0.5 and 1. Formally, this means
\begin{align*}
d_i=
\begin{cases}
0.5 & \text{if } d_i^\ast \leq 0.5\\
d_i^\ast&\text{if }d_i^\ast\in(0.5,q_{\text{emp},0.98})\\
1&\text{if } d_i^\ast\geq q_{\text{emp},0.98},
\end{cases} 
\end{align*}
where $d_i^\ast=\Phi_{\mu_1,\lambda_1}(\xi_{i1})$ and where $q_{\text{emp},0.98}$ denotes the empirical 98 percent quantile of the sample $\{d_1^\ast,\dots,d_n^\ast\}$. The structure of the Dep.-Dis.~DGP resembles our application and represents a quite challenging example in contrast to DGPs with bigger masses at 1. For the Ind.-Dis.~DGP we draw $d_i$ from an adjusted Bernoulli (Ber) distribution on $\{0.5,1\}$ with parameter $p=0.5$, and for the Ind.-Con.~DGP we draw $d_i$ from a standard uniform distribution.

For each DGP, we generate 500 replications of $n = 50, 150, 250$, and $500$ functions $X_i(t)$ evaluated at equidistant grid points $t_1 ,\ldots,t_p$ in $[0,1]$ with $p=501$, where all values $X_i(t_j)$ with $t_j>d_i$ are considered missing and replaced by NAs. Our simulation is implemented using the statistical language \textsf{R} \citep{rcore}, where we make use of the package \texttt{fda} \citep{fda} to create the Fourier system. In order to maintain the sin-cosine pairs of the Fourier system, we consider only odd numbers of basis functions $J=3,5, \ldots, J_{\text{max}}$ with $J_{\text{max}}=51$.
\begin{table}[!ht]
\centering
\tabcolsep=0.20cm
\caption{Integrated squared bias and variance.}
\vspace{0.08cm}
\label{tab:sim}
\begin{tabular}{cl rr c rr c rr c rr}
\toprule
&& \multicolumn{2}{c}{Dep.-Dis.} && \multicolumn{2}{c}{Dep.-Con.} && \multicolumn{2}{c}{Ind.-Dis.} && \multicolumn{2}{c}{Ind.-Con.} \\
\cline{3-4}\cline{6-7}\cline{9-10}\cline{12-13}\\[-2ex]
n&                               & Bias & Var. && Bias & Var. && Bias & Var. && Bias & Var.\\
\midrule
$\phantom{5}50$
&$\widehat \mu_{\text{FTC}}$    & 0.0  & 1.0  && 0.0  & 1.0  && 0.0  & 0.9  && 0.0  & 1.0  \\
&$\widehat \mu$                 & 3.2  & 0.8  && 6.0  & 1.3  && 0.0  & 0.9  && 0.1  & 1.9  \\
$150$
&$\widehat \mu_{\text{FTC}}$    & 0.0  & 0.3  && 0.0  & 0.4  && 0.0  & 0.3  && 0.0  & 0.3  \\
& $\widehat \mu$                & 3.2  & 0.3  && 6.8  & 0.5  && 0.0  & 0.3  && 0.0  & 0.6  \\
$250$
&$\widehat \mu_{\text{FTC}}$    & 0.0  & 0.2  && 0.0  & 0.2  && 0.0  & 0.2  && 0.0  & 0.2  \\
& $\widehat \mu$                & 3.1  & 0.2  && 6.7  & 0.3  && 0.0  & 0.2  && 0.0  & 0.3  \\
$500$
&$\widehat \mu_{\text{FTC}}$    & 0.0  & 0.1  && 0.0  & 0.1  && 0.0  & 0.1  && 0.0  & 0.1  \\
&$\widehat \mu$                 & 3.2  & 0.1  && 6.7  & 0.1  && 0.0  & 0.1  && 0.0  & 0.2  \\
\midrule
$\phantom{5}50$
&$\widehat \sigma_{\text{FTC}}$ &  0.1  & 40.7  &&  0.5  & 47.1  && 0.4  & 43.8  && 0.7  & 44.1  \\
&$\widehat \sigma$              & 31.3  & 27.1  && 43.2  & 38.4  && 0.3  & 36.7  && 7.6  & 57.6  \\
$150$
&$\widehat \sigma_{\text{FTC}}$ &  0.1  & 13.3  &&  0.1  & 15.7  && 0.1  & 15.1  && 0.1  & 15.0  \\
&$\widehat \sigma$              & 30.4  &  8.7  && 38.0  & 16.6  && 0.0  & 12.2  && 0.7  & 28.3  \\
$250$
&$\widehat \sigma_{\text{FTC}}$ &  0.1  &  7.9  &&  0.1  &  9.2  && 0.1  &  9.4  && 0.1  &  8.8  \\
&$\widehat \sigma$              & 30.7  &  5.4  && 38.7  & 10.2  && 0.0  &  7.5  && 0.4  & 17.7  \\
$500$
&$\widehat \sigma_{\text{FTC}}$ &  0.1  &  4.3  &&  0.0  &  4.8  && 0.1  &  4.7  && 0.1  &  4.6  \\
&$\widehat \sigma$              &  0.1  &  4.3  &&  0.0  &  4.8  && 0.1  &  4.7  && 0.1  &  8.7  \\
\bottomrule
\end{tabular}
\end{table}

Table \ref{tab:sim} contains our simulation results for our FTC-estimators $\widehat\mu_{\operatorname{FTC}}$ and $\widehat\sigma_{\operatorname{FTC}}$. We report the (doubly) integrated squared bias $(\int_0^1)\int_0^1 \operatorname{Bias}^2\left(\cdot\right)$ and variance $(\int_0^1)\int_0^1 \operatorname{Var}\left(\cdot\right)$ for $\widehat\mu_{\operatorname{FTC}}$ ($\widehat\sigma_{\operatorname{FTC}}$) and compare them with the estimation errors of the classical estimators $\widehat\mu$ ($\widehat\sigma$). The results clearly demonstrate a systematic bias of the classical estimators $\widehat\mu$ and $\widehat \sigma$ under the Dep.~DGPs (see the Dep.-Dis.~and Dep.-Con.~columns), which is non-vanishing as $n$ increases. By contrast, our FTC-estimators show essentially no bias---despite Violation (V) of the MCAR assumption. If the MCAR assumption holds, both estimators lead to equivalent results up to some minor numerical issues; the slightly negative performance of the FTC-estimator $\widehat{\sigma}_{\text{FTC}}$ in the Ind.-Dis.~DGP is due to computing numerical integrals and the disadvantage of the classical estimators in the Ind.-Con.~DGP is due to the influence of $\xi_1$ with the largest variance $\lambda_1$ on the outer domains in $[d_{\operatorname{min}},1]$, where only a small percentage of the sample is observed.

To demonstrate the performance of our practical approach for identifying Violation (V), we present simulation results for the adapted multiple testing procedure proposed by \citet{RomWol2005b}; see Table \ref{tab:sim2}. We use a significance level of $\alpha = 0.05$ and report the percentages of replications falling under the scenarios Null, (V), and Other (see Section \ref{sec:vio}). The simulation results provide a convincing picture of our practical identification procedure. The procedure complies with the strong control of the FWER in all DGPs with $n\geq 150$, where the true scenario (V) for the Dep.~DGPs and the true scenario Null for the Ind.~DGPs are only rejected in approximately 5\% of the replications. 
For $n=50$ the procedure shows a very poor performance, since selecting a basis system with up to $J_{\text{max}}=51$ basis elements in such a small sample typically leads to misspecified basis choices resulting in unstable test results. However, the test results stabilize when using a reduced number of $J_{\text{max}}=31$ basis elements for $n=50$; see second row in Table \ref{tab:sim2}. From our experience with this method, one should apply this procedure with $J_{\text{max}}=51$ for sample sizes $n\geq 150$, but a reduced number of basis elements $31\leq J_{\text{max}}<51$ for $50\leq n<150$.
\begin{table}[!ht]
\centering
\tabcolsep=0.17cm
\caption{Selection errors for Null, (V) and Other (in \%) -- $J_{\text{max}}=51$.}
\vspace{0.08cm}
\label{tab:sim2}
\begin{tabular}{c ccc c ccc c ccc c ccc}
\toprule
& \multicolumn{3}{c}{Dep.-Dis.} && \multicolumn{3}{c}{Dep.-Con.} && \multicolumn{3}{c}{Ind.-Dis.} && \multicolumn{3}{c}{Ind.-Con.}\\
\cline{2-4}\cline{6-8}\cline{10-12}\cline{14-16}\\[-2ex]
 n& {\small Null} &{\small (V)} &{\small Other} &&{\small Null} &{\small (V)} &{\small Other} &&{\small Null} &{\small (V) }&{\small Other }&&{\small Null} &{\small (V) }&{\small Other} \\
\midrule
$\phantom{5}50\phantom{^\ast}$&          85.0&15.0&0.0&&49.6          &50.4&0.0&&100.0          &0.0&0.0&&100.0          &0.0&0.0\\
$\phantom{5}50^\ast$          &\phantom{8}0.0&97.2&2.8&&\phantom{4}2.0&95.0&3.0&&\phantom{1}93.8&1.2&5.0&&\phantom{1}98.6&0.2&1.2\\
$150\phantom{^\ast}$          &\phantom{8}0.0&99.0&1.0&&\phantom{4}0.0&97.8&2.2&&\phantom{1}98.2&0.2&1.6&&\phantom{1}97.6&0.2&2.2\\
$250\phantom{^\ast}$          &\phantom{8}0.0&96.6&3.4&&\phantom{4}0.0&99.0&1.0&&\phantom{1}97.4&0.4&2.2&&\phantom{1}97.6&0.6&1.8\\
$500\phantom{^\ast}$          &\phantom{8}0.0&95.6&4.4&&\phantom{4}0.0&94.6&5.4&&\phantom{1}96.0&0.6&3.4&&\phantom{1}95.2&0.2&4.6\\
\bottomrule
\multicolumn{16}{l}{$^\ast$ $J_{\text{max}}=31$.}
\end{tabular}
\end{table}

\section{German Control Reserve Market Study} \label{sec:emp}
In this section, we use the above described estimation procedure to estimate the mean price curve of control power prices from the German Control Reserve Market. The German Control Reserve Market has an important role in maintaining grid stability, which is one of the basic responsibilities of electricity markets. An effective tool to guarantee grid stability are reserve capacities. Therefore, the German Federal Network Agency (FNA), a public institution, determines each week a sufficiently large amount of reserve capacity $d_i$ that has to be bought, by law, by the Grid Control Cooperation\footnote{The GCC is a merger of the four German Transport System Operators (TSOs) \url{www.50hertz.com}, \url{www.amprion.net}, \url{www.transnetbw.de}, and \url{www.tennettso.de}.} (GCC). The purpose of the German Control Reserve Market is to generate  market prices at which the GCC has to buy the dictated reserve capacities from the electricity providers. 

\begin{figure}[!ht]
\includegraphics[width=.49\textwidth]{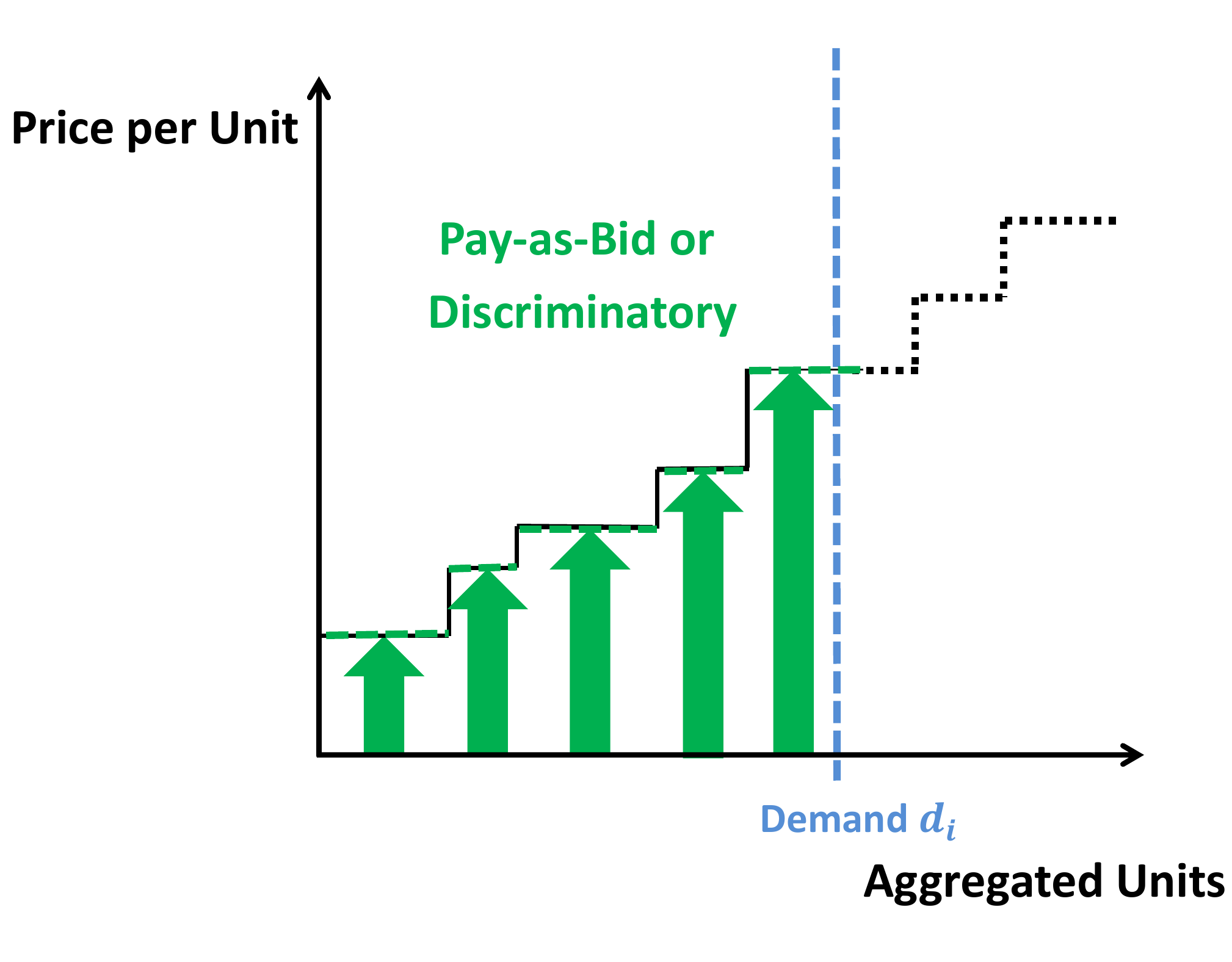}\hfill
\includegraphics[width=.49\textwidth]{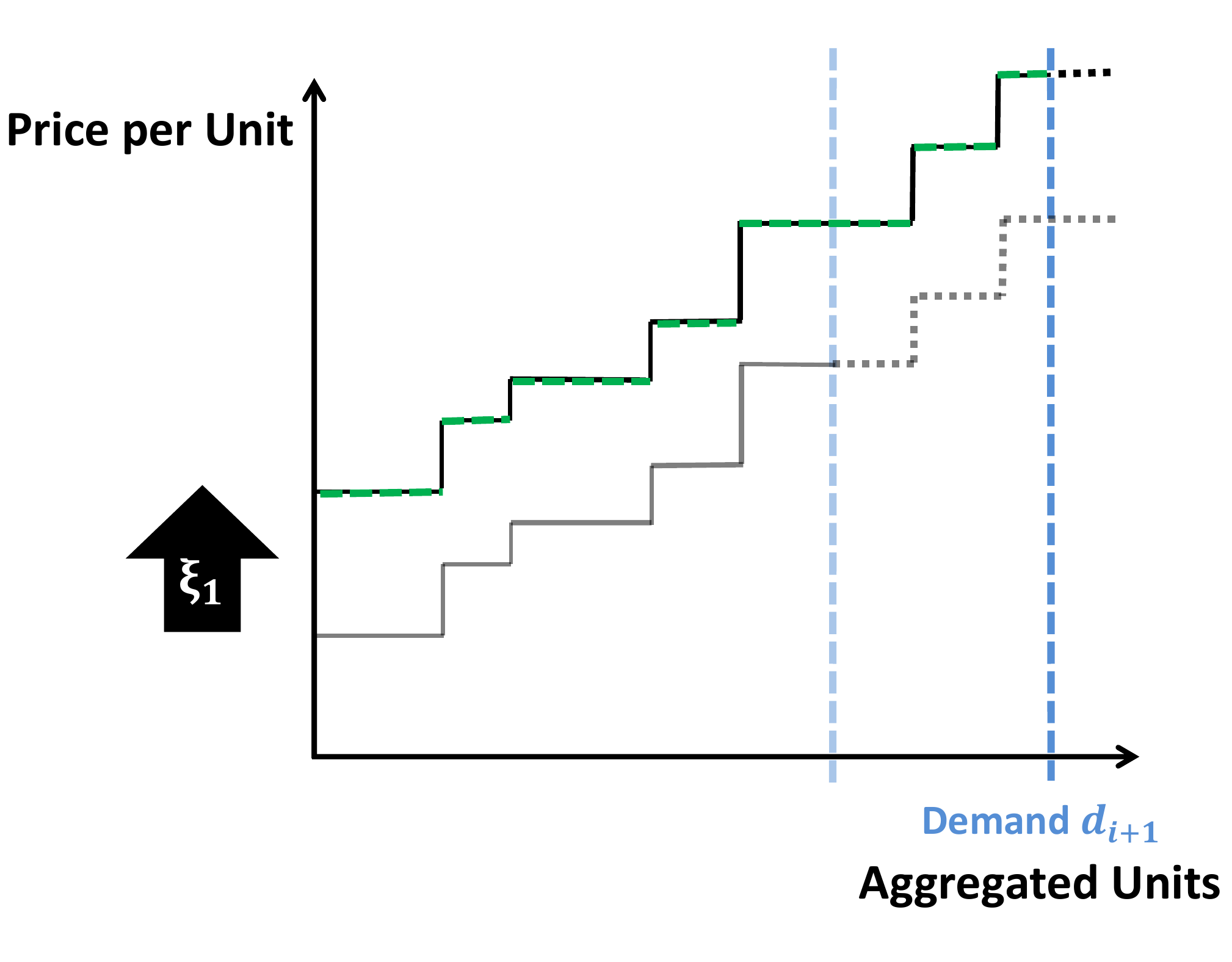}
\caption{Left: The Pay-as-bid pricing mechanism, where each successful bidder receives the price of his bid. Right: A vertical shift dependency of the price curves due to changes in the announced amount of reserve capacity $d_i$.}
\label{fig:auc}
\end{figure}
\begin{figure}[!hbt]
\includegraphics[width=1\textwidth]{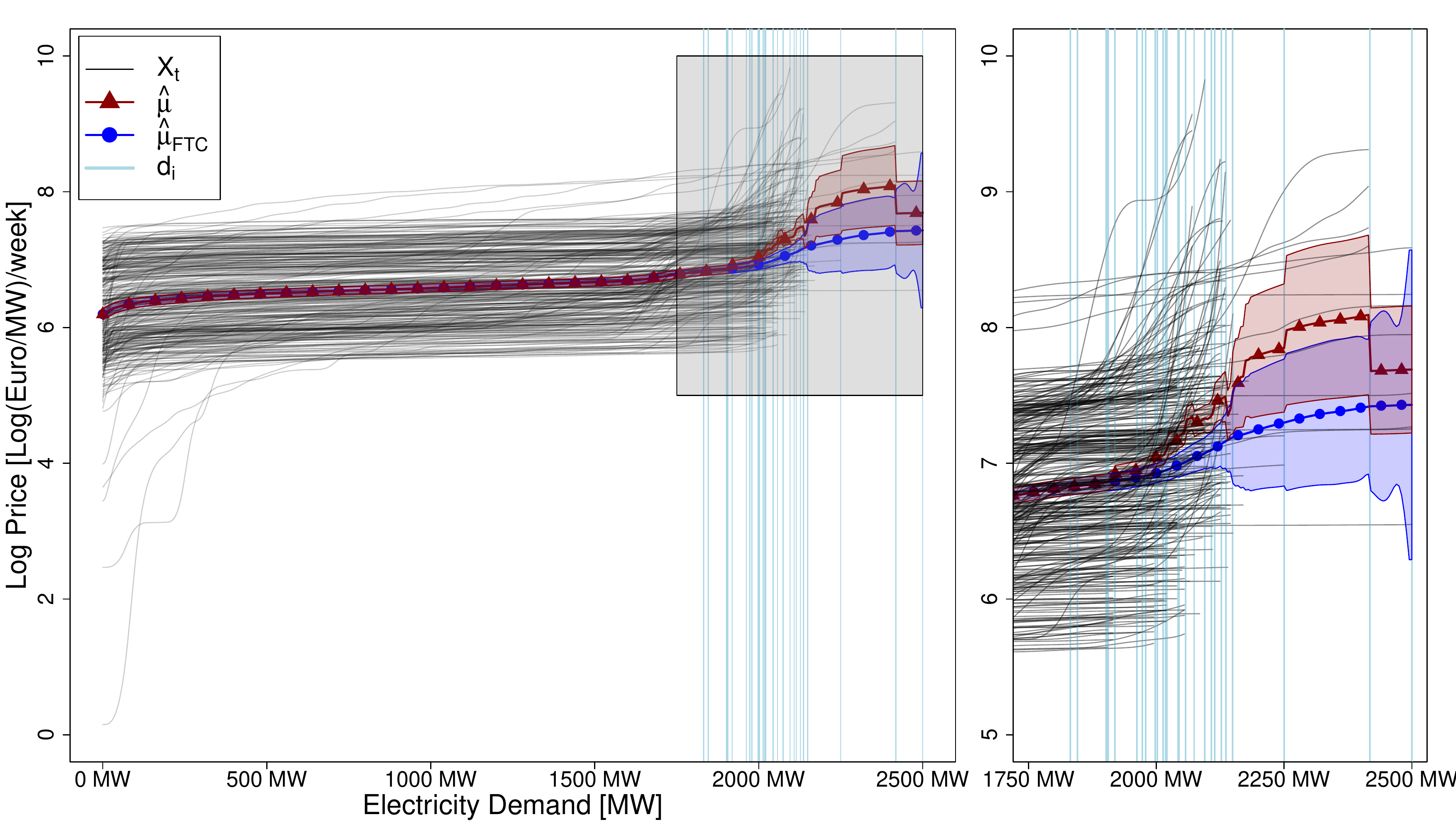}
\caption{Classical mean (solid with triangles) and FTC mean estimates (solid with circles) computed from $n=337$ partially observed price curves.}
\label{fig:ftc_mean2}
\end{figure}

The German Control Reserve Market uses a so-called ``pay-as-bid'' auction, where every successful bidder (i.e., the electricity providers) receives his own (sealed offered) price at which the bidder is willing to provide a certain amount of electricity. In this market design it is optimal to bid the (unknown) maximum price the GCC is eventually forced to pay (see, e.g., \citeauthor{GriOckZo2008}, \citeyear{GriOckZo2008}, or \citeauthor{EngGri2009}, \citeyear{EngGri2009}). Therefore, all bidders try to predict this unknown maximum price and the best predictor is the publicly announced amount of reserve capacity $d_i$. If the FNA dictates a large value of $d_i$, bidders tend to increase their prices, which leads to vertical shifts in the price curve and vice versa. That is, price curves $X_i$ that are observed over larger sub-domains $[0,d_i]$ tend to have overall higher price levels and vice versa. This relationship is shown schematically in the right plot of Figure \ref{fig:auc} and can be clearly seen in the data (Figure \ref{fig:ftc_mean2}). This relatively simple dependency between the sub-domains $[0,d_i]$ and the level-shifts in the price curves $X_i$ motivated the above described Violation (V) of the MCAR assumption.  

\begin{figure}[!ht]
\begin{center}
\includegraphics[width=.9\textwidth]{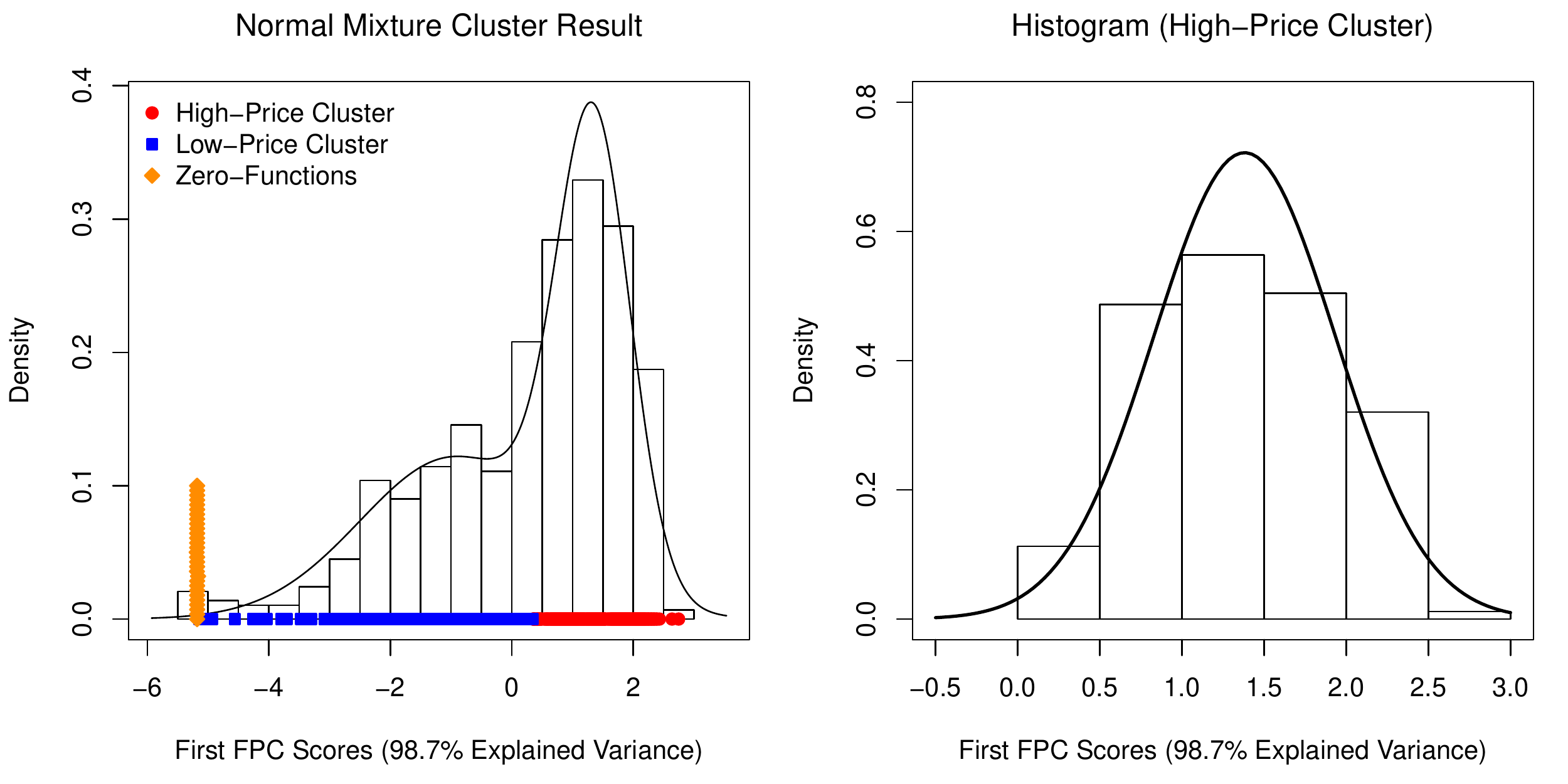} 
\caption{Distribution of the first principal component scores of the log-price curves computed with respect to the fully observed interval $[a,d_{\operatorname{min}}]$ of the original sample.}
\label{fig:clust}  
\end{center}   
\end{figure}
\noindent\textbf{Data, Preprocessing, and Exploratory Analysis:} The data are freely available from \url{www.regelleistung.net}, and we provide the data set as well as implementations of our estimators in the accompanying \textsf{R}-package \texttt{PartiallyFD}. The weekly data covers the time horizon from June 27th, 2011, to April, 17th, 2017. Two special auctions (January 2nd, 2012 and December 31st, 2012) resulted in extreme prices ($\geq$ 20.000 EUR/MW) and are removed. In order to account for some very large prices in the data, we analyze logarithmized price curves, where all prices lower than 1 EUR/MW are mapped to 1 EUR/MW. To obtain a functional data sample, we pre-smooth each logarithmized price curve using cubic monotone P-splines in order to keep the monotonicity and the differentiability of the bid curves. The necessary function \texttt{monotSpline} is included in the accompanying \textsf{R}-package \texttt{PartiallyFD}. All functions are observed over the initial interval $[a,d_{\operatorname{min}}]=[0\hspace{1mm}\text{MW},1832\hspace{1mm}\text{MW}]$, but are only partially observed over the final interval $(d_{\operatorname{min}},b]=(1832\hspace{1mm}\text{MW},2500\hspace{1mm}\text{MW}]$. As in our simulation study, we evaluate the curves at $p=501$ equidistant points $t_1 = 0, t_2 = 5, \ldots, t_{501}=2500$ over the complete domain $[a,b] = [0\hspace{1mm}\text{MW}, 2500\hspace{1mm}\text{MW}]$; evaluation points within the missing sub-domains $[d_i,b]$ are filled by NAs. The log-price curves are relatively simply structured random functions and their first Functional Principal Component (FPC), computed with respect to the fully observed lower interval, accounts for $98.7\%$ of the total variance. (See, for instance, \citeauthor{RamSil2005}, \citeyear{RamSil2005}, Ch.~8, for an introduction to FPC Analysis). This allows for an exploratory data analysis with respect to the first FPC scores which reveals that the original sample of log-price curves is bi-modal and has an additional point-mass at zero-price functions. Therefore, we remove the zero-price functions from the sample and perform a normal mixture cluster analysis on the first FPC scores using the \textsf{R}-package \texttt{mclust} of \cite{FraRaf2002}. The result of this exploratory data analysis is shown in Figure \ref{fig:clust}. For our further analysis we focus on the practically relevant high-price cluster with $n=337$ price curves. Normality is not necessary for our estimators to be consistent under Violation (V), but is needed in our practical test procedure described in Section \ref{sec:vio}. Standard normality tests reject the null hypothesis of normality due to the hard thresholding used to allocate the data into the single clusters; however, the empirical distribution of the high-price cluster is not substantially far from a normal distribution (see Figure \ref{fig:clust}).  

\smallskip

We test the functional data sample for Violation (V) as described in Section \ref{sec:vio} where the procedure leads to scenario (V), i.e., the identification of Violation (V). For robustness, we additionally checked $J_{\text{max}}=51\pm 10$, which also lead to Violation (V). To estimate the mean function, we use our FTC-estimator $\widehat{\mu}_{\text{FTC}}$ as defined in \eqref{mean_ftc} and compare the estimation result with that from using the classical estimator $\widehat\mu$. The difference between the classical estimator $\widehat \mu(t)$, which is inconsistent under Violation (V), and our FTC-estimator $\widehat \mu_{\text{FTC}}(t)$ is obvious: the systematically missing parts of the curves lead to a positive bias in $\widehat\mu(t)$ resulting in an implausible, non-monotonous, non-smooth mean curve. In contrast, our FTC-estimator leads to a smooth and monotonous mean curve, which is perfectly plausible given the monotonicity of the price curves. Nonetheless, the confidence intervals are overlapping, but this is not surprising, given the sparseness of the data at the critical upper ending of the domain. The bias in the classical mean estimator $\widehat{\mu}$ results also in a defective estimate of the covariance function when using the classical covariance estimator $\widehat{\sigma}$. A plot of the covariance surface is omitted for reasons of space, but can be produced using the \textsf{R}-scripts of the online supplementary materials.

\section*{Acknowledgements}
The authors thank Prof.~Rolf Tschernig (University of Regensburg) for his valuable suggestions which helped to improve this research work. Many thanks go also to two anonymous referees for their constructive comments which had a very positive impact on this paper too.

\appendix

\section{Appendix}\label{sec:app}

\subsection{Methodology}\label{ssec:met}
\begin{proposition}\label{lem:bias}
Under (B) and under Violation $(V)$, the estimator $\widehat\mu(t)$ is biased, i.e., $\E{\widehat\mu(t)} = \mu(t) + \Delta(t)$, where $\Delta(t) = \mathbb{E}\left[\mathbb{E}\left[\xi_{i1}\hspace{1mm}|\hspace{1mm}O_i(t)\right]\right]-\mu_1\psi_1(t)$.
\end{proposition}
\noindent\textbf{Proof of Proposition \ref{lem:bias}:}
\begin{align*}
\mathbb{E}\left[\widehat\mu(t) \right]  =& \mathbb{E} \left[\mathbb{E}\left[\widehat\mu(t)\hspace{1mm} | \hspace{1mm}O_1(t), \ldots, O_n(t) \right]  \right]\\
 =& \mathbb{E} \left[\frac{I_1(t)}{\sum_{i=1}^n O_i(t)} \sum_{i=1}^n O_i(t)  \left(\mathbb{E}\left[\xi_{i1}\hspace{1mm}|\hspace{1mm}O_i(t)\right] \psi_1(t) + \sum_{j\geq 2}  \mu_{j}\psi_j(t)\right)\right]  \\
 =& \mathbb{E} \left[\frac{I_1(t)}{\sum_{i=1}^n O_i(t)} \sum_{i=1}^n O_i(t)  \sum_{j\geq 2}  \mu_{j}\psi_j(t) \right] + \mathbb{E} \left[\frac{I_1(t)}{\sum_{i=1}^n O_i(t)} \sum_{i=1}^n O_i(t)  \mathbb{E}\left[\xi_{i1}\hspace{1mm}|\hspace{1mm}O_i(t)\right] \psi_1(t)\right] 
\end{align*}
The first summand deviates from $\mu(t)$ only by $\mu_1 \psi_1(t)$:
\begin{align*}
\mathbb{E} \left[\frac{I_1(t)}{\sum_{i=1}^n O_i(t)} \sum_{i=1}^n O_i(t)  \sum_{j\geq 2}  \mu_{j}\psi_j(t) \right] =& \mathbb{E} \left[\frac{I_1(t)}{\sum_{i=1}^n O_i(t)} \sum_{i=1}^n O_i(t)  \left(\sum_{j\geq 1}  \mu_{j}\psi_j(t) - \mu_1 \psi_1(t)\right)\right]   \\
=& \mu(t) - \mu_1 \psi_1(t)
\end{align*}
The second part deviates from $\mu_1 \psi_1(t)$ by $\Delta(t)$ defined via 
\begin{align*}
\Delta(t)=\mathbb{E}\left[ \frac{I_1(t)}{\sum_{i=1}^n O_i(t)} \sum_{i=1}^n O_i(t)\mathbb{E}\left[\xi_{i1}\hspace{1mm}|\hspace{1mm}O_i(t)\right]\right]-\mu_1\psi_1(t)
\end{align*}
which results in
\begin{align*}
\mathbb{E} \left[\frac{I_1(t)}{\sum_{i=1}^n O_i(t)} \sum_{i=1}^n O_i(t)  \mathbb{E}\left[\xi_{i1}\hspace{1mm}|\hspace{1mm}O_i(t)\right] \psi_1(t)\right]=& \mathbb{E} \left[\frac{I_1(t)}{\sum_{i=1}^n O_i(t)} \sum_{i=1}^n O_i(t) \left(\mu_1\psi_1(t) + \Delta(t)\right) \right]  \\
=& \mu_1\psi_1(t) + \Delta(t)
\end{align*}
\noindent The bias of $\widehat \sigma (s,t)$ can be derived in a similar manner.

\bigskip

\noindent\textbf{Proof of Theorem \ref{Th1}:} \label{proof}
First, we focus on $\widehat{\mu}_{\operatorname{FTC}}(t)$. When $t\in[0,d_{\operatorname{min}}]$, we observe $p(t)=1$, i.e., 100\% of the sample. Therefore, the MCAR assumption is not violated and $\widehat \mu$ is unbiased and consistent. 

For $t \in (d_{\operatorname{min}}, 1]$, the estimator is defined as
\begin{align*}
\widehat\mu_{\text{FTC}}(t)
&=\int_{d_{\operatorname{min}}}^t \widehat\mu^{(1)}(z) dz + \widehat\mu(d_{\operatorname{min}})
\end{align*}
by decomposing $\widehat{\mu}(t)$ via the FTC. A direct application of the product rule delivers the representation of 
\begin{align}
\widehat{\mu}^{(1)}(z)&=\frac{J(z)}{\sum_{i=1}^n O_i(z)} \sum_{i=1}^n X^{(1)}_i(z)O_i(z)dz \label{eq:ftc}
\end{align}
as the mean estimator of the first derivatives $X^{(1)}_i(t)$. 

Under our setup with $\psi_1 \equiv 1$ and under Violation (V), the derivatives $X_{i}^{(1)}(z)=\sum_{j\geq 2}\xi_{ij} \psi_j^{(1)}(z)$ are independent from $O_i(z)$ such that there is no violation of the MCAR when estimating $\mu^{(1)}(z)$ by $\widehat\mu^{(1)}(z)$ for $z\in[d_{\operatorname{min}},1]$. It follows by the Weak Law of Large Numbers that $\widehat\mu^{(1)}(z)\to_p\mu^{(1)}(z)$ as $n\to\infty$. For the second summand in \eqref{mean_ftc}---since $p(d_{\operatorname{min}})=1$---we see that $n^{-1}\sum_{i=1}^n X_i(d_{\operatorname{min}})\to_p\mu(d_{\operatorname{min}})$ as $n\to\infty$. With the Continuous Mapping Theoreom (CMT) we finally deduce the unbiasedness and consistency of $\widehat{\mu}_{\operatorname{FTC}}(t)$ for all $t \in[0,1]$. This reasoning is later used to show consistency in the case of $\widehat{\sigma}_{\text{FTC}}(s,t)$ as well.

In the case of $\widehat{\sigma}_{\operatorname{FTC}}(s,t)$, we go stepwise through all combinations for $s,t \in [0, d_{\operatorname{min}}] \cup (d_{\operatorname{min}},1]$.   
\bit
\item[\textbf{I:}] For $s,t \in [0, d_{\operatorname{min}}]$. \\
Here, $p(t) = 1$, i.e., there is no missing data and the standard estimator $\widehat{\sigma}(s,t)$ as in \eqref{var_kra} is consistent and therefore $\widehat{\sigma}_{\operatorname{FTC}}(s,t)$ as well.

\item[\textbf{II:}] Without loss of generality, we assume $s\in [0, d_{\operatorname{min}}]$ and $t \in (d_{\operatorname{min}}, 1]$.\\
Then, the estimator is defined as 
\beqs
\widehat{\sigma}_{\text{FTC}}(s,t) = \int_{d_{\operatorname{min}}}^t \widehat{\sigma}^{(0,1)}(s, z_2)dz_2 + \widehat{ \sigma}(s,d_{\operatorname{min}}),
\eeqs
where $\widehat{\sigma}^{(k,\ell)}(s,t)=(\partial^{k +\ell}/(\partial s^{k}\partial t^{\ell}))\widehat{\sigma}(s,t)$ with $k,\ell\in\N$. Hence, using the same arguments as after \eqref{eq:ftc}, applied to 
\beqs
\widehat{\sigma}^{(0,1)}(s, z_2) = \frac{I_2(s,z_2)}{\sum_{i=1}^n U_i(s,z_2)} \sum_{i=1}^n U_i(s,z_2) \Big[X_i(s) - \widehat\mu(s)\Big] \Big[X^{(1)}_i(z_2) - \widehat\mu^{(1)}(z_2)\Big]
\eeqs
and to $\widehat{ \sigma}(s,d_{\operatorname{min}})$, 
we can deduce the consistency of $\widehat{\sigma}^{(0,1)}(s,t)$ and $\widehat{ \sigma}(s,d_{\operatorname{min}})$ for $s\in [0, d_{\operatorname{min}}]$ and $z_2 \in (d_{\operatorname{min}}, 1]$. \\
Parallel reasoning holds for the case where $s\in (d_{\operatorname{min}}, 1]$ and $t \in [0, d_{\operatorname{min}}]$.
 
\item[\textbf{III:}] For $s,t \in (d_{\operatorname{min}}, 1]$. \\
In order to derive the estimator for the case where $s,t \in (d_{\operatorname{min}}, 1]$, observe that the following three equations hold due to the FTC:
\begin{align}
\sigma(s,t)   &= \int_{d_{\operatorname{min}}}^t \frac{\partial}{\partial z_2}\sigma(s, z_2)dz_2 + \sigma(s,d_{\operatorname{min}})\label{eq:III_1}\\
\sigma(s,z_2) &= \int_{d_{\operatorname{min}}}^s \frac{\partial}{\partial z_1}\sigma(z_1, z_2)dz_1 + \sigma(d_{\operatorname{min}},z_2)\label{eq:III_2}\\
\int_{d_{\operatorname{min}}}^t \frac{\partial}{\partial z_2}\sigma(d_{\operatorname{min}},z_2)dz_2&=\sigma(d_{\operatorname{min}},t)-\sigma(d_{\operatorname{min}},d_{\operatorname{min}})\label{eq:III_3}
\end{align}
Plugging \eqref{eq:III_2} into \eqref{eq:III_1} and using the linearity of the operations involved yields
\begin{align}
\sigma(s,t)  &= \int_{d_{\operatorname{min}}}^t \frac{\partial}{\partial z_2} \left(\int_{d_{\operatorname{min}}}^s \frac{\partial}{\partial z_1}\sigma(z_1, z_2)dz_1 + \sigma(d_{\operatorname{min}},z_2)\right)dz_2 + \sigma(s,d_{\operatorname{min}})\nonumber\\
\Leftrightarrow
\sigma(s,t)  &= \int_{d_{\operatorname{min}}}^t \int_{d_{\operatorname{min}}}^s \frac{\partial^2}{\partial z_1\partial z_2}\sigma(z_1, z_2)dz_1dz_2 + \int_{d_{\operatorname{min}}}^t \frac{\partial}{\partial z_2}\sigma(d_{\operatorname{min}},z_2)dz_2 + \sigma(s,d_{\operatorname{min}}).
\label{eq:III_4}
\end{align}
Plugging \eqref{eq:III_3} into \eqref{eq:III_4} yields
\begin{align}
\sigma(s,t)  &= \int_{d_{\operatorname{min}}}^t \int_{d_{\operatorname{min}}}^s \frac{\partial^2}{\partial z_1\partial z_2}\sigma(z_1, z_2)dz_1dz_2 + \sigma(d_{\operatorname{min}},t) + \sigma(s,d_{\operatorname{min}})-\sigma(d_{\operatorname{min}},d_{\operatorname{min}}).\nonumber
\end{align}
Replacing $\sigma(d_{\operatorname{min}},t)$ with a corresponding version of \eqref{eq:III_1} and $\sigma(s,d_{\operatorname{min}})$ with a corresponding version of \eqref{eq:III_2} yields
\begin{align*}
\sigma(s,t)  &= \int_{d_{\operatorname{min}}}^t \int_{d_{\operatorname{min}}}^s \frac{\partial^2}{\partial z_1\partial z_2}\sigma(z_1, z_2)dz_1dz_2\\
& + \int_{d_{\operatorname{min}}}^t \frac{\partial}{\partial z_2}\sigma(d_{\operatorname{min}}, z_2)dz_2 + \sigma(d_{\operatorname{min}},d_{\operatorname{min}})\\
& + \int_{d_{\operatorname{min}}}^s \frac{\partial}{\partial z_1}\sigma(z_1, d_{\operatorname{min}})dz_1 + \sigma(d_{\operatorname{min}},d_{\operatorname{min}})\\
& -\sigma(d_{\operatorname{min}},d_{\operatorname{min}}).
\end{align*}
Simplifying and using the notation of Definition \ref{def:FTC_cov} leads to:
\begin{align*}
\sigma(s,t) &= \int_{d_{\operatorname{min}}}^t \int_{d_{\operatorname{min}}}^s \sigma^{(1,1)}(z_1, z_2)dz_1dz_2\\
& + \int_{d_{\operatorname{min}}}^t \sigma^{(0,1)}(d_{\operatorname{min}}, z_2)dz_2 
  + \int_{d_{\operatorname{min}}}^s \sigma^{(1,0)}(z_1, d_{\operatorname{min}})dz_1 
  + \sigma(d_{\operatorname{min}},d_{\operatorname{min}}),
\end{align*}
where, following the equivalent arguments as used above, $\sigma^{(1,1)}(z_1, z_2)$, $\sigma^{(0,1)}(d_{\operatorname{min}}, z_2)$, $\sigma^{(1,0)}(z_1, d_{\operatorname{min}})$, and $\sigma(d_{\operatorname{min}},d_{\operatorname{min}})$ can be estimated consistently using
\begin{align*}
\widehat{\sigma}^{(\ell,k)}(s,t)=
\frac{I_2(s,t)}{\sum_{i=1}^n U_i(s,t)} \sum_{i=1}^n U_i(s,t) \left[X_i^{(\ell)}(s) - \widehat{\mu}^{(\ell)}(s) \right] \left[X_i^{(k)}(t) - \widehat{\mu}^{(k)}(t) \right],\quad \ell,k\in\{0,1\}.
\end{align*}
\eit

\bibliographystyle{Chicago}
\bibliography{literatur}

\begin{thebibliography}{}

\bibitem[\protect\citeauthoryear{Delaigle and Hall}{Delaigle and
  Hall}{2013}]{DelHal2013}
Delaigle, A. and P.~Hall (2013).
\newblock Classification using censored functional data.
\newblock {\em Journal of the American Statistical Association\/}~{\em
  108\/}(504), 1269--1283.

\bibitem[\protect\citeauthoryear{Delaigle and Hall}{Delaigle and
  Hall}{2016}]{DelHal2016}
Delaigle, A. and P.~Hall (2016).
\newblock Approximating fragmented functional data by segments of {Markov}
  chains.
\newblock {\em Biometrika\/}~{\em 103\/}(4), 779--799.

\bibitem[\protect\citeauthoryear{Engelmann and Grimm}{Engelmann and
  Grimm}{2009}]{EngGri2009}
Engelmann, D. and V.~Grimm (2009).
\newblock Bidding behaviour in multi-unit auctions--an experimental
  investigation.
\newblock {\em The Economic Journal\/}~{\em 119\/}(537), 855--882.

\bibitem[\protect\citeauthoryear{Ferraty and Vieu}{Ferraty and
  Vieu}{2006}]{FerVie2006}
Ferraty, F. and P.~Vieu (2006).
\newblock {\em Nonparametric Functional Data Analysis - Theory and Practice}.
\newblock Spinger.

\bibitem[\protect\citeauthoryear{Fraley and Raftery}{Fraley and
  Raftery}{2002}]{FraRaf2002}
Fraley, C. and A.~E. Raftery (2002).
\newblock Model-based clustering, discriminant analysis, and density
  estimation.
\newblock {\em Journal of the American Statistical Association\/}~{\em
  97\/}(458), 611--631.

\bibitem[\protect\citeauthoryear{Gellar, Colantuoni, Needham, and
  Crainiceanu}{Gellar et~al.}{2014}]{GelColNeeCrai2014}
Gellar, J.~E., E.~Colantuoni, D.~M. Needham, and C.~M. Crainiceanu (2014).
\newblock Variable-domain functional regression for modeling {ICU} data.
\newblock {\em Journal of the American Statistical Association\/}~{\em
  109\/}(508), 1425--1439.

\bibitem[\protect\citeauthoryear{Goldberg, Ritov, and Mandelbaum}{Goldberg
  et~al.}{2014}]{GolRitMan2014}
Goldberg, Y., Y.~Ritov, and A.~Mandelbaum (2014).
\newblock Predicting the continuation of a function with applications to call
  center data.
\newblock {\em Journal of Statistical Planning and Inference\/}~{\em 147},
  53--65.

\bibitem[\protect\citeauthoryear{Grimm, Ockenfels, and Zoettl}{Grimm
  et~al.}{2008}]{GriOckZo2008}
Grimm, V., A.~Ockenfels, and G.~Zoettl (2008).
\newblock Strommarktdesign: Zur {A}usgestaltung der {A}uktionsregeln an der
  {EEX}.
\newblock {\em Zeitschrift für Energiewirtschaft\/}~(3), 147--161.

\bibitem[\protect\citeauthoryear{Gromenko, Kokoszka, and Sojka}{Gromenko
  et~al.}{2017}]{GroKokSoj2017}
Gromenko, O., P.~Kokoszka, and J.~Sojka (2017).
\newblock Evaluation of the cooling trend in the ionosphere using functional
  regression with incomplete curves.
\newblock {\em The Annals of Applied Statistics\/}~{\em 11\/}(2), 898--918.

\bibitem[\protect\citeauthoryear{Horv{\'a}th and Kokoszka}{Horv{\'a}th and
  Kokoszka}{2012}]{HorKok2012}
Horv{\'a}th, L. and P.~Kokoszka (2012).
\newblock {\em Inference for Functional Data with Applications}.
\newblock Springer.

\bibitem[\protect\citeauthoryear{Hsing and Eubank}{Hsing and
  Eubank}{2015}]{HsiEub2015}
Hsing, T. and R.~Eubank (2015).
\newblock {\em Theoretical Foundations of Functional Data Analysis, with an
  Introduction to Linear Operators}.
\newblock John Wiley \& Sons.

\bibitem[\protect\citeauthoryear{James, Hastie, and Sugar}{James
  et~al.}{2000}]{JamHasSu2000}
James, G.~M., T.~J. Hastie, and C.~A. Sugar (2000).
\newblock Principal component models for sparse functional data.
\newblock {\em Biometrika\/}~{\em 87\/}(3), 587--602.

\bibitem[\protect\citeauthoryear{James and Sugar}{James and
  Sugar}{2003}]{JamSug2003}
James, G.~M. and C.~A. Sugar (2003).
\newblock Clustering for sparsely sampled functional data.
\newblock {\em Journal of the American Statistical Association\/}~{\em
  98\/}(462), 397--408.

\bibitem[\protect\citeauthoryear{Kraus}{Kraus}{2015}]{Kra2015}
Kraus, D. (2015).
\newblock Components and completion of partially observed functional data.
\newblock {\em Journal of the Royal Statistical Society\/}~{\em 77\/}(4),
  777--801.

\bibitem[\protect\citeauthoryear{Liebl}{Liebl}{2013}]{Lie2013}
Liebl, D. (2013).
\newblock Modeling and forecasting electricity spot prices: A functional data
  perspective.
\newblock {\em The Annals of Applied Statistics\/}~{\em 7\/}(3), 1562--1592.

\bibitem[\protect\citeauthoryear{Liebl and Kneip}{Liebl and
  Kneip}{2018}]{LieKne2018}
Liebl, D. and A.~Kneip (2018).
\newblock {\em On the optimal reconstruction of partially observed functional
  data}.
\newblock arXiv:1710.10099.

\bibitem[\protect\citeauthoryear{Little and Rubin}{Little and
  Rubin}{2014}]{LitRub2014}
Little, R.~J. and D.~B. Rubin (2014).
\newblock {\em Statistical analysis with missing data}.
\newblock John Wiley \& Sons.

\bibitem[\protect\citeauthoryear{Ramsay and Silverman}{Ramsay and
  Silverman}{2005}]{RamSil2005}
Ramsay, J. and B.~Silverman (2005).
\newblock {\em Functional Data Analysis\/} (2. ed.).
\newblock Springer.

\bibitem[\protect\citeauthoryear{Ramsay, Wickham, Graves, and Hooker}{Ramsay
  et~al.}{2014}]{fda}
Ramsay, J.~O., H.~Wickham, S.~Graves, and G.~Hooker (2014).
\newblock {\em fda: Functional Data Analysis}.
\newblock \textsf{R} package version 2.4.4.

\bibitem[\protect\citeauthoryear{{\textsf{R} Core Team}}{{\textsf{R} Core
  Team}}{2017}]{rcore}
{\textsf{R} Core Team} (2017).
\newblock {\em \textsf{R}: A Language and Environment for Statistical
  Computing}.
\newblock Vienna, Austria: R Foundation for Statistical Computing.

\bibitem[\protect\citeauthoryear{Romano and Wolf}{Romano and
  Wolf}{2005}]{RomWol2005b}
Romano, J.~P. and M.~Wolf (2005).
\newblock Exact and approximate stepdown methods for multiple hypothesis
  testing.
\newblock {\em Journal of the American Statistical Association\/}~{\em
  100\/}(469), 94--108.

\bibitem[\protect\citeauthoryear{Yao, M{\"u}ller, and Wang}{Yao
  et~al.}{2005a}]{YaoMueWa2005}
Yao, F., H.-G. M{\"u}ller, and J.-L. Wang (2005a).
\newblock Functional data analysis for sparse longitudinal data.
\newblock {\em Journal of the American Statistical Association\/}~{\em
  100\/}(470), 577--590.

\bibitem[\protect\citeauthoryear{Yao, M{\"u}ller, and Wang}{Yao
  et~al.}{2005b}]{YaoMueWan2005}
Yao, F., H.-G. M{\"u}ller, and J.-L. Wang (2005b).
\newblock Functional linear regression analysis for longitudinal data.
\newblock {\em The Annals of Statistics\/}~{\em 33\/}(6), 2873--2903.

\bibitem[\protect\citeauthoryear{Zhou, Serban, Gebraeel, and M{\"u}ller}{Zhou
  et~al.}{2014}]{ZhoSerGebMue2014}
Zhou, R.~R., N.~Serban, N.~Gebraeel, and H.-G. M{\"u}ller (2014).
\newblock A functional time warping approach to modeling and monitoring
  truncated degradation signals.
\newblock {\em Technometrics\/}~{\em 56\/}(1), 67--77.

\end{thebibliography}
\end{document}